\documentclass[twocolumn,prb,superscriptaddress,nofootinbib]{revtex4-2}
\usepackage[english]{babel}
\usepackage{boldline,multirow,xcolor,colortbl}
\usepackage{times}
\usepackage{graphicx}
\usepackage{graphics}
\usepackage{amsmath}
\usepackage{amsfonts}
\usepackage{amssymb}
\usepackage{amsbsy}
\usepackage{dsfont}
\usepackage{epstopdf}
\usepackage{makeidx}
\usepackage{subfigure}
\usepackage{color} 
\usepackage{pgf}
\usepackage{bm}
\usepackage{tikz} 
\usepackage[normalem]{ulem}
\usepackage[symbol]{footmisc}
\usepackage{float}
\usepackage{placeins}
\usepackage{hyperref}
\newcommand{\be}{\begin{equation}}
\newcommand{\ee}{\end{equation}}
\newcommand{\bea}{\begin{eqnarray}}
\newcommand{\eea}{\end{eqnarray}}

\begin{document}
\title{Entangled two-plasmon generation in carbon nanotubes and graphene coated wires}


\author{Y. Muniz}
\email{yurimuniz@pos.if.ufrj.br}
\affiliation{Instituto de F\'{\i}sica, Universidade Federal do Rio de Janeiro, Caixa Postal 68528, Rio de Janeiro 21941-972, RJ, Brazil}

\author{P. P. Abrantes}
\email{patricia@pos.if.ufrj.br}
\affiliation{Instituto de F\'{\i}sica, Universidade Federal do Rio de Janeiro, Caixa Postal 68528, Rio de Janeiro 21941-972, RJ, Brazil}

\author{L. Mart\'in-Moreno}
\email{lmm@unizar.es}
\affiliation{Instituto de Ciencia de Materiales de Arag\'on and Departamento de F\'{\i}sica de la Materia Condensada, CSIC-Universidad de Zaragoza, E-50009 Zaragoza, Spain}
\affiliation{Center for Photonics and 2D Materials, Moscow Institute of Physics and Technology, Dolgoprudny 141700, Russia}

\author{F. A. Pinheiro}
\email{fpinheiro@if.ufrj.br}
\affiliation{Instituto de F\'{\i}sica, Universidade Federal do Rio de Janeiro, Caixa Postal 68528, Rio de Janeiro 21941-972, RJ, Brazil}

\author{C. Farina}
\email{farina@if.ufrj.br}
\affiliation{Instituto de F\'{\i}sica, Universidade Federal do Rio de Janeiro, Caixa Postal 68528, Rio de Janeiro 21941-972, RJ, Brazil}

\author{W. J. M. Kort-Kamp}
\email{kortkamp@lanl.gov}
\affiliation{Theoretical Division, Los Alamos National Laboratory, MS B262, Los Alamos, New Mexico 87545, United States}	

\begin{abstract}
We investigate the two-plasmon spontaneous decay of a quantum emitter near single-walled carbon nanotubes (SWCNT) and graphene-coated wires (GCWs). We demonstrate efficient, enhanced generation of two-plasmon entangled states in SWCNTs due to the strong coupling between tunable guided plasmons and the quantum emitter. We predict two-plasmon emission rates more than twelve orders of magnitude higher than in free-space, with average lifetimes of a few dozens of nanoseconds. Given their low dimensionality, these systems could be more efficient for generating and detecting entangled plasmons in comparison to extended graphene. Indeed, we achieve tunable spectrum of emission in GCWs, where sharp resonances occur precisely at the plasmons' minimum excitation frequencies. We show that, by changing the material properties of the GCW's dielectric core, one could tailor the dominant modes and frequencies of the emitted entangled plasmons while keeping the decay rate ten orders of magnitude higher than in free-space.  By unveiling the unique properties of two-plasmon spontaneous emission processes in the presence of low dimensional carbon-based nanomaterials, our findings set the basis for a novel material platform with applications to on-chip quantum information technologies.
\end{abstract}

\maketitle

Generating and manipulating non-classical states of light is of pivotal importance in nanophotonics, quantum technologies, and cryptography~\cite{o2009photonic}. In order to produce single photons on-demand for various applications such as imaging and quantum sensing, atomic~\cite{haroche2013}, solid state platforms~\cite{aharonovich2016}, and entangled photon pairs in nonlinear crystals~\cite{kwiat95} are typically employed. Two-photon spontaneous emission (TPSE) is an alternative approach for generating entangled photon pairs~\cite{goppert1931, zalialiutdinov2018}, which has been achieved in different scenarios such as atomic~\cite{lipeles1965, bannett1982, cesar1996} and semiconductor systems~\cite{hayat2008}, and biexciton-exciton decay in quantum dots~\cite{ota2011}. In the TPSE quantum process, an excited emitter decays to its ground state by simultaneously emitting a pair of entangled photons~\cite{wang2019}. TPSE processes have much broader emission spectra in comparison to single-photon ones since any combination of photon energies satisfying the constraint of total energy conservation is allowed. However, being a second-order process in the emitter-field coupling constant, TPSE results in an emission rate that is typically several orders of magnitude slower than one-photon emission one, typically restricting its applicability in quantum technologies. 

Progress in the fields of plasmonics and metamaterials have allowed for enhancing TPSE, and hence expanding its versatility and applicability~\cite{hayat2007,poddubny2012}. Indeed, an impressive increase of the TPSE rate can be achieved by engineering the local density of optical states (LDOS) in nanostructured electromagnetic environments, such as planar photonic systems~\cite{hayat2007,muniz2020}, optical cavities~\cite{goncalves2020}, phonon-polaritons dielectric systems~\cite{rivera2017}, and aperiodic bandgap structures~\cite{luca2021}. For instance, plasmon-assisted collective TPSE has been observed in bulk semiconductors coupled to nanoantenna arrays~\cite{nevet2010} with reduced radiative emission enhancement. Remarkably, by tailoring the LDOS in plasmonic and polaritonic sub-wavelength electromagnetic nanostructures one may achieve TPSE rates that are orders of magnitude larger than one-photon spontaneous emission rates~\cite{rivera2017}. Very recently, atomically thin and two-dimensional plasmonic systems have emerged as another versatile material platform to harness TPSE processes from single emitters. Indeed, atomically thin nanostructures can be effectively employed to control TPSE, resulting in giant far-field two-photon production, enabling tailored photonic and plasmonic entangled states, and plasmon-assisted single-photon creation orders of magnitude larger than standard one-photon emission~\cite{muniz2020}. In addition, spontaneous decay into two-plasmon polaritons in graphene monolayers is predicted to be more than ten orders of magnitude larger than two-photon transitions~\cite{rivera2016}.

Here we introduce another carbon-based nanomaterial platform to tailor TPSE processes, namely single-walled carbon nanotubes (SWCNT) and graphene-coated wires (GCWs). Specifically, we investigate the two-plasmon spontaneous decay of a quantum emitter near SWCNT and GCWs to demonstrate that the strong coupling between tunable guided plasmons and the emitter allows for efficient generation of two-plasmon entangled states. This material platform also enables unprecedented control over spectral lineshapes of emission due to the coupling with different plasmonic modes. We also find that, in contrast to other plasmonic systems, GCWs allow one to tune the TPSE spectrum without significant variations in the emitter's lifetime. For this reason, GCWs enable one to independently customize plasmon emission rate and frequency distribution.

\section{Model and theory}
\begin{figure*}
\begin{center}
\includegraphics[width=1\linewidth,keepaspectratio]{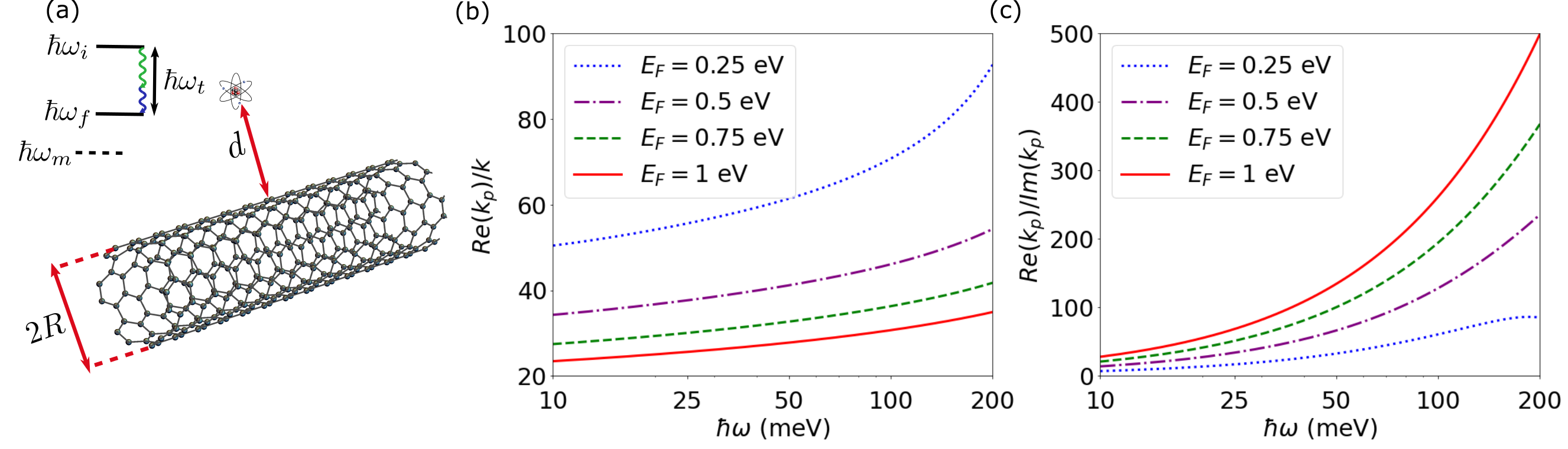}
\caption{{(a)} Schematics of the system under study. A quantum emitter separated by a distance $d$ from a SWCNT of radius $R$. {(b)} Real part of the plasmon's wavevector $k_p$ for the fundamental mode ($n = 0$) normalized by the free-space wavevector $k = \omega/c$ as a function of the light oscillation frequency for different values of the Fermi energy. {(c)} The ratio between the real and imaginary parts of $k_p$ of the fundamental plasmonic mode versus the free-space frequency for different values of the Fermi energy.}
\label{fig1}
\end{center}
\end{figure*}
Consider a quantum emitter separated by a distance $d$ from an infinitely long nanowire of radius $R$. The surface of the wire is graphene while its inner region may be empty (as for the case of a SWCNT) or filled with a dielectric medium of relative electric permittivity $\varepsilon$ (as for a GCW). The emitter is initially in a spherically symmetric state of energy $\hbar\omega_i$ and can decay to a final one (with the same symmetry) of energy $\hbar\omega_f$ via two-quanta spontaneous emission assisted by intermediate states of energy $\hbar\omega_m$ (Fig. 1a). Since both initial and final states have the same parity, no one-quantum transitions between them are allowed due to selection rules. However, we note that the multilevel character of the quantum emitter implies the existence of other one- and two-quanta competing decay pathways. The rotational and translational symmetries with respect to the nanowire axis allow for the diagonalization of the imaginary part of the electromagnetic Green's tensor in the frequency-independent basis $\{\bm{\hat{\rho}}, \bm{\hat{z}}, \bm{\hat{\phi}}\}$ (the basis vectors in cylindrical coordinates). In this case, and assuming that the emitter has its size $l_e$ much smaller than the transition wavelengths, the corresponding TPSE rate can be written as~\cite{muniz2019}
\begin{equation}
\Gamma(d)\! =\!\!  \int_0^{\omega_t}\!\!\!\!\!\!\! d\omega \frac{\gamma_0(\omega)}{3}\!\!\sum_{a}\!
P_a(d,\omega)P_a(d,\omega_t-\omega),\!
\label{GammaTotal}
\end{equation}
where $\omega_t = \omega_i-\omega_f$ is the transition frequency, $P_a(d,\omega)$ is the single-photon Purcell factor for a transition dipole moment oriented along the direction of the unit vector ${\bf \hat{e}}_a \in \{\bm{\hat{\rho}}, \bm{\hat{z}}, \bm{\hat{\phi}}\}$, and $\gamma_0(\omega)\sim \omega^3(\omega_t-\omega)^3|\mathbb{D}(\omega,\omega_t - \omega)|^2l_e^6/c^6 $ is the free-space TPSE spectral density. The tensor $\mathbb{D}$ encodes the emitter's electronic structure and is given by
\begin{equation}
\mathbb{D}(\omega_1,\omega_2) = \sum_m \left[\frac{{\bf d}_{im}{\bf d}_{mf}}{\omega_{im} - \omega_1} + \frac{{\bf d}_{mf}{\bf d}_{im}}{\omega_{im} - \omega_2}\right],\label{D}
\end{equation}
where ${\bf d}_{ab}$ is the emitter $a \rightarrow b$ transition dipole moment and the summation must be made over all intermediate quantum states. The TPSE spectrum $\gamma(d,\omega)$ is defined as the integrand of Eq. \eqref{GammaTotal} and is always symmetric with respect to $\omega_t/2$ due to energy conservation.

The Purcell factors are proportional to the local density of states of the considered photonic environment~\cite{carminati2015} and can be obtained by calculating the electric field of an oscillating dipole near the wire using the relation \cite{novotny2012} 
\begin{equation}
P_a(d, \omega) = 1 + \frac{6\pi \varepsilon_0c^3}{p_a^2\omega^3}\mathrm{Im}\{{\bf p}_a \cdot {\bf E}_R(d,\omega)\}, 
\end{equation}
where ${\bf p}_a = p_a\hat{{\bf e}}_a$ is the dipole moment, and ${\bf E}({\bf r}) = {\bf E}_0({\bf r}) + {\bf E}_R({\bf r})$ for $\rho > R$ and ${\bf E}({\bf r}) = {\bf E}_T({\bf r})$ for $\rho < R$. Here, ${\bf E}_0$ is the free-space field generated by the dipole, and ${\bf E}_R$ and ${\bf E}_T$ are the fields reflected and transmitted by the nanostructure, respectively. We consider transitions with wavelengths much larger than the geometrical parameters of the system, i.e., $\lambda \gg R, d$. In this case, it is sufficient to calculate the field in the quasi-static approximation \cite{luis2015}. In this regime, $ {\bf E}({\bf r}) = -\nabla({\bf p}_a \cdot \nabla) \Phi({\bf r})$, where $\Phi$ is the scalar electrostatic potential, which satisfies the Poisson equation $\nabla^2 \Phi({\bf r}) = \delta({\bf r} - d\bm{\hat{\rho}})/\varepsilon_0$ and the boundary conditions regarding the continuity of the parallel component of the electric field, $[{\bf E}_0(R) + {\bf E}_R(R) - {\bf E}_T(R)]\times \bm{\hat{\rho}} = {\bf 0}$, and the discontinuity of its perpendicular component due to the induced surface charge density, $[{\bf E}_0(R) + {\bf E}_R(R) - \varepsilon{\bf E}_T(R)]\cdot \bm{\hat{\rho}} = \rho_{\text{ind}}/\varepsilon_0$. The induced surface charge can be obtained from the continuity equation, which yields $i\omega\rho_{\text{ind}} = \sigma\nabla\cdot{\bf E}_\parallel({\bf r})$, where $\sigma$ is the graphene conductivity. Here we assume that $R$ is sufficiently large ($ \gtrsim 1$ nm) so that we can neglect finite-size and chirality effects of the nanotube~\cite{deabajo2014}. By expanding the potential in cylindrical coordinates and applying the boundary conditions one finds
\begin{align}
\Phi_0({\bf r}) &= \sum_{m = 0}^{\infty}\frac{b_m\cos(m\theta)}{2\pi^2\varepsilon_0}\!\!\int_0^\infty\!\!\!\!\!\! dk\cos(kz)I_m(k\rho)K_m(k\rho_d),
\\
\Phi_R({\bf r}) &= \sum_{m = 0}^{\infty}\frac{b_m\cos(m\theta)}{2\pi^2\varepsilon_0}\!\!\int_0^\infty\!\!\!\!\!\! dk\, r_m\cos(kz)K_m(k\rho)K_m(k\rho_d),
\\
\Phi_T({\bf r}) &= \sum_{m = 0}^{\infty}\frac{b_m\cos(m\theta)}{2\pi^2\varepsilon_0}\!\!\int_0^\infty\!\!\!\!\!\! dk\, t_m\cos(kz)I_m(k\rho)K_m(k\rho_d),
\end{align}
where $b_m = 2 - \delta_{m0}$, $\rho_d = R + d$, $I_m$ and $K_m$ are the modified Bessel functions of the first and second kind, respectively, and the reflection ($r_m$) and transmission ($t_m$) Fresnel coefficients are given by
\begin{align}
r_m &= -\frac{(\varepsilon - 1)I_mI_m'kR + \Delta_mI_m^2}{(\varepsilon I_m'K_m - I_mK_m')kR + \Delta_mI_mK_m}, \label{rm}
\\
t_m &= 1 + \frac{K_m}{I_m}r_m,
\end{align}
with $\Delta_m = (i\sigma/\varepsilon_0\omega R)(m^2 + k^2R^2)$, the Bessel functions being evaluated at $kR$, and the prime represents derivative with respect to the argument. Finally, we can use the previous expression for the reflected scalar potential to calculate the reflected electric field and obtain the Purcell factors for each relevant dipolar orientation as follows
\begin{align}
P_\rho(d, \omega) &= 1 - \sum_{m = 0}^{\infty}\frac{3c^3b_m}{\pi\omega^3}\int_0^\infty\!\!\!\!\!\! dk\, k^2\mathrm{Im}\{r_m\}[K'_m(k\rho_d)]^2,
\\
P_z(d, \omega) &= 1 - \sum_{m = 0}^{\infty}\frac{3c^3b_m}{\pi\omega^3}\int_0^\infty\!\!\!\!\!\! dk\, k^2\mathrm{Im}\{r_m\}K^2_m(k\rho_d),
\\
P_\phi(d, \omega) &= 1 - \sum_{m = 1}^{\infty}\frac{6c^3m^2}{\pi\rho_d^2\omega^3}\int_0^\infty\!\!\!\!\!\! dk\, \mathrm{Im}\{r_m\}K^2_m(k\rho_d).
\end{align}
It is important to mention that the previous equations apply to both SWCNT and GCW cases considered in this paper, with SWCNTs being the particular case of $\varepsilon = 1$. For a SWCNT, we can reobtain the simpler expression of the reflection coefficient derived in \cite{luis2015} by using the Wronskian identity $I_m'(x)K_m(x) - I_m(x)K_m'(x) = 1/x$.

The plasmon dispersion relations are given by the poles of the Fresnel coefficients, which can be obtained by solving the following transcendental equation,
\begin{equation}
\frac{(m^2 + k^2R^2)I_mK_m}{(\varepsilon I_m'K_m - I_mK_m')kR} = \frac{i\varepsilon_0\omega R}{\sigma(\omega)}.
\end{equation}
When we consider a Drude model for the graphene conductivity, i.e., $\sigma(\omega) = ie^2E_F/\pi\hbar^2(\omega + i/\tau)$ where $E_F$ is the Fermi energy and $\tau$ is the relaxation time, in the limit of small dissipation ($\tau \rightarrow \infty$), the free-space oscillation frequency can be directly expressed in terms of the plasmon wavevector as
\begin{equation}
\hbar\omega_m(k) = \sqrt{\frac{e^2E_F}{\pi\varepsilon_0 R}\times\frac{(m^2 + k^2R^2)I_mK_m}{(\varepsilon I_m'K_m - I_mK_m')kR}}.
\end{equation}
This equation gives the energy required to excite the plasmonic mode $m$ with propagation wavevector $k$. Since each guided plasmon can be supported regardless of the value of $k$, by taking the limit $k \rightarrow 0$ we get the minimum amount of energy required to excite the plasmonic mode $m$. Using the appropriate Taylor expansions for the modified Bessel functions and taking this limit, we find 
\begin{equation}
\hbar\omega_m^{(\text{min})} = \sqrt{\frac{e^2E_Fm}{(1 + \varepsilon)\pi\varepsilon_0 R}}. \label{omega_min}
\end{equation}
From the previous equation, we conclude that the fundamental mode ($m = 0$) can be excited at any oscillation frequency, while other modes require some amount of energy to exist. Such a difference can be explained by their non-trivial angular profile~\cite{cuevas2015}, which imposes a constraint over the plasmon's wavelength $\lambda_g^{\phi}$ across the $\phi$-direction, namely, $m\lambda_g^{\phi} = 2\pi R$. This can be demonstrated by inserting $k_g = 2\pi/\lambda_g^{\phi} = m/R$ into the plasmon's dispersion relation of the extended graphene in the quasi-static approximation, given by $\hbar\omega_g = \sqrt{e^2E_Fk_g/(1 + \varepsilon)\pi\varepsilon_0}$~\cite{bludov2013}. As a consequence of Eq. \eqref{omega_min}, depending on the geometric and material properties of the wire, different plasmonic modes contribute to the TPSE spectrum of emission. In short, every mode with $\omega_m^{(\text{min})} < \omega_t$ contributes to the spectrum, which can be tuned by modifying the system's properties such as the wire radius, the Fermi energy, or even the relative permittivity by a proper choice of the inner dielectric medium.

\section{Results}
\subsection{TPSE in single-walled carbon nanotube}

In this section, we consider the case of an emitter near a SWCNT ($\varepsilon = 1$), shown in Fig 1a. SWCNTs typically have diameters in the range of a few nanometers~\cite{ganesh2013}, which imposes constraints regarding the appearance of non-fundamental plasmonic modes in the TPSE spectrum. Indeed, let us consider a large nanotube of radius $R = 5$ nm, which is at the limit of what can be achieved with state-of-the-art nanofabrication techniques~\cite{ma2009}. From Eq. \eqref{omega_min} the minimum excitation energy of the 1st-order mode for $E_F = 500$ meV is given by $\hbar\omega^{(\text{min})} \sim 537$ meV, which may induce interband transitions in the nanotube instead of exciting plasmons. Even if the Fermi energy is increased to $1$ eV, the first mode only exists above $760$ meV, which is near the limit of the mid-infrared spectral range where plasmons have been shown to exist for graphene. Any other mode has an excitation frequency in the regime dominated by interband transitions for any value of the Fermi energy and, consequently, would not show up in the TPSE spectral lineshapes. For this reason, the fundamental mode dominates the TPSE spectrum of an emitter near a SWCNT. In all subsequent discussion, we consider a SWCNT with $2$ nm of radius and electron mobility of $10^4$ cm$^2$V$^{-1}$s$^{-1}$, which has been previously demonstrated in graphene samples \cite{bolotin2008,dean2010}. In Fig. 1b we show the light confinement of the fundamental plasmonic mode for frequencies below $200$ meV. In this regime, we notice higher confinements for smaller values of the chemical potential, which leads to higher spontaneous emission enhancements due to the direct impact of confinement on the local density of states~\cite{rivera2016}. On the other hand, Fig. 1c shows the ratio between the real and imaginary components of the propagation wavevector, which is typically employed as the figure of merit (FOM) to characterize the relationship between a plasmon's wavelength and its propagation length in the SWCNT. Note that this FOM increases with the Fermi energy, implying a trade-off between light confinement and plasmon propagation in SWCNTs. Nevertheless, the fundamental plasmonic mode still offers strong confinements and long propagation lengths in this frequency regime. 

In Fig. 2a we plot the TPSE rate normalized by the free-space two-photon decay rate as a function of the distance between the emitter and the nanotube surface. As an example, we consider the $6s \rightarrow 5s$ transition in hydrogen ($\hbar\omega_t \sim 166$ meV), and carry the summation in Eq. \eqref{D} over the 20 first intermediate states ($2p$ to $20p$ states), obtaining satisfactory convergence of results. We notice an extreme enhancement, $\sim 10^{12}$, of the TPSE rate in the near-field regime at $d = 10$ nm and higher at smaller distances. We also note that the dependence of the emitter's two-plasmon emission rate on the distance to the SWCNT surface presents a noticeable change of behavior for $d \sim 100$ nm. This contrasts with the one-plasmon emission rate~\cite{luis2015}, which approximately follows an exponential decay with the distance~\cite{zeng2016}. This difference occurs because the TPSE rate involves an integral over a broad spectrum of frequencies below $\omega_t$, and the exponential coefficient of $P(d,\omega)$ is not a constant. In the upper right inset we plot the spectral density as a function of distance for three frequencies of emission. We notice that $\gamma$ goes to zero faster at frequencies close to $\omega_t/2$. Furthermore, exactly at $\omega_t/2$ there is no change of behavior in the far field since $\gamma(d,\omega_t/2) \sim P^2(d,\omega_t/2)$, which follows an exponential decay with the distance. The corresponding spectral lineshapes are shown in the lower left inset of Fig 2a. For any value of the Fermi energy, there exists a similar broadband spectrum where no particular frequency of emission is favoured. This is a consequence of the fact that the TPSE is only affected by the fundamental plasmonic mode of the SWCNT. In Fig. 2b we plot the TPSE quantum efficiency (QE), which is defined as
\begin{figure}[t!]
\begin{center}
\includegraphics[width=0.98\linewidth]{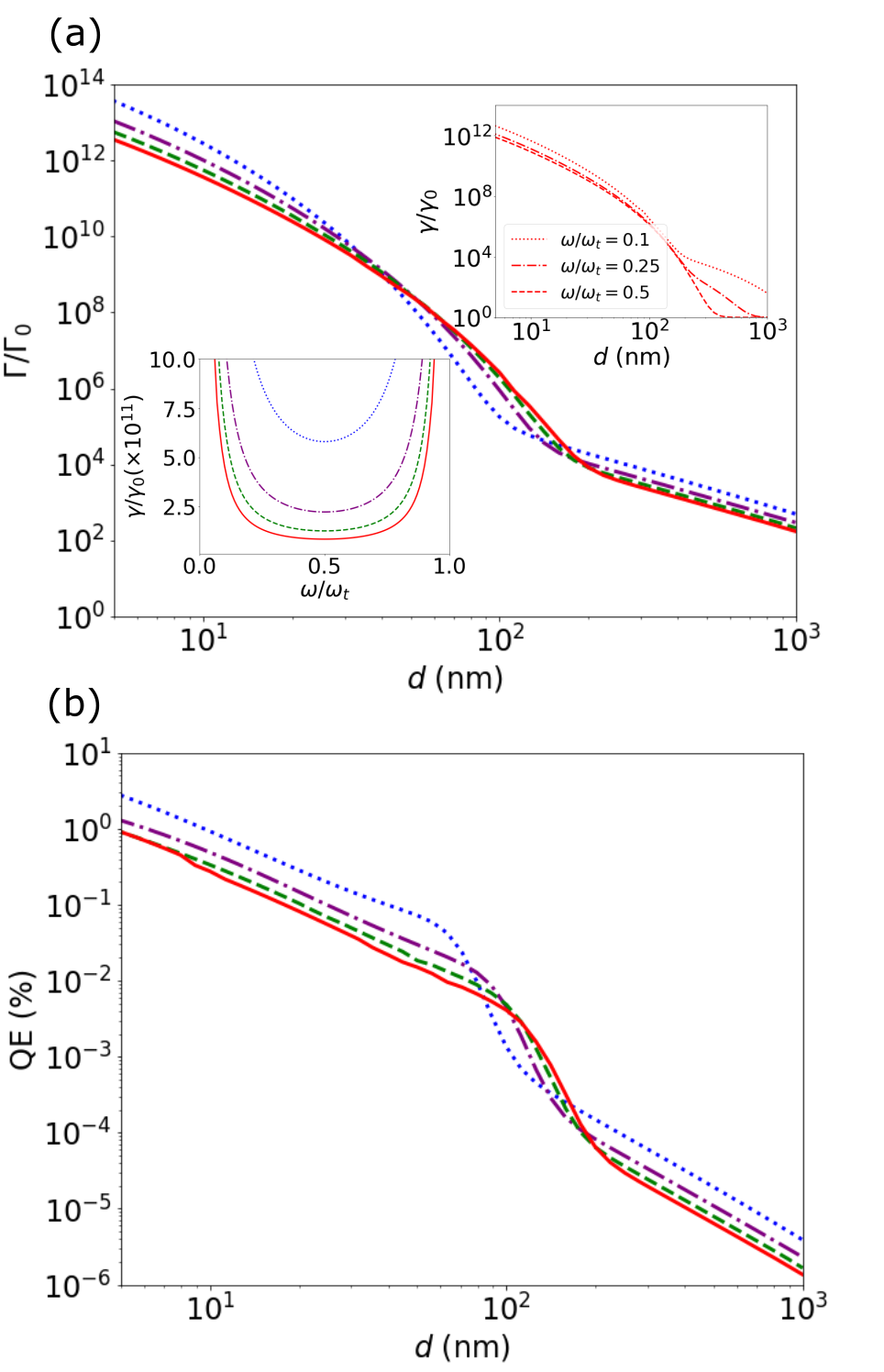}
\caption{(a)  Normalized TPSE rate for the $6s \rightarrow 5s$ transition in hydrogen ($\hbar\omega_t \approx 166$ meV) as a function of the distance between the emitter and the surface of the SWCNT. Upper right inset: Normalized TPSE spectral density as a function of distance for three frequencies of emission. In this inset, the Fermi energy is given by $E_F = 1$ eV. Lower left inset: Normalized TPSE spectral density at $d = 10$ nm. Since only the fundamental plasmonic mode is present, we observe a broadband spectrum of emission. The divergences at $\omega = 0$ and $\omega = \omega_t$ are solely due to the normalization by the free-space spectral density, which goes to zero at the boundaries of the spectrum. (b) Quantum efficiency given by Eq. \eqref{QE} of TPSE for the $6s \rightarrow 5s$ transition in hydrogen as a function of distance. In both plots, $E_F = \{0.25, 0.5, 0.75, 1\}$ eV (dotted blue, dash-dotted purple, dashed green, and solid red lines, respectively)}
\label{fig2}
\end{center}
\end{figure}
\begin{equation}
\text{QE} = \frac{\Gamma}{\Gamma + \Gamma^{1\text{q}}_{6s \rightarrow 5p} + \Gamma^{1\text{q}}_{6s \rightarrow 4p} + \Gamma^{1\text{q}}_{6s \rightarrow 3p} + \Gamma^{1\text{q}}_{6s \rightarrow 2p}}, \label{QE}
\end{equation}
where $\Gamma^{1\text{q}}_{6s \rightarrow np}$ ($n = 5, 4, 3, 2$) is the first-order spontaneous emission rate between the $6s$ and the $np$ state.
\begin{figure*}
\begin{center}
\includegraphics[width=1\linewidth,keepaspectratio]{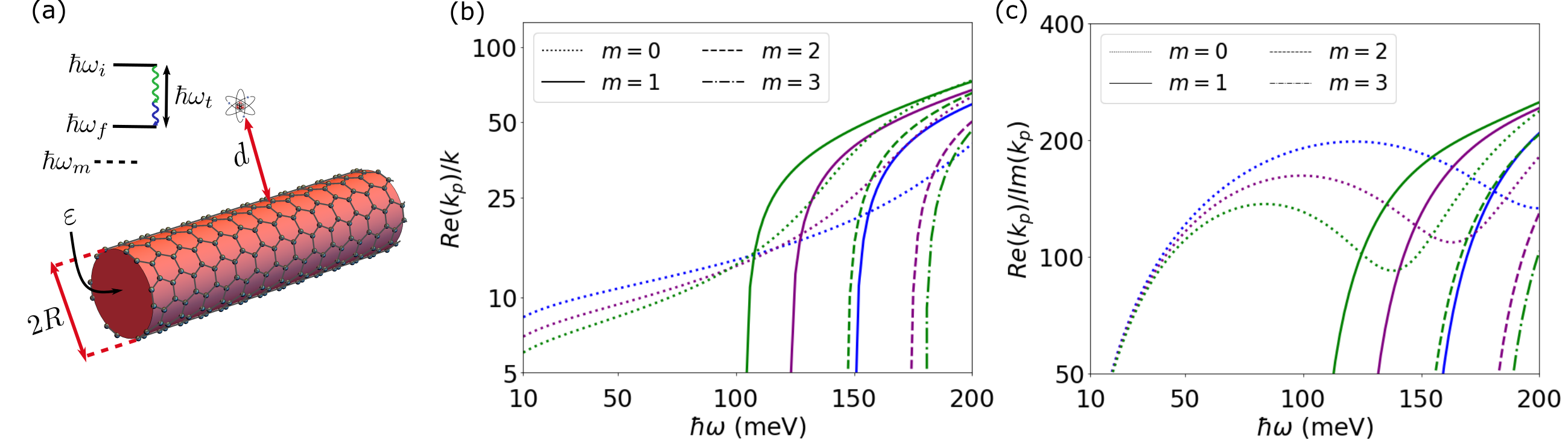}
\caption{{(a)} Schematics of the system: a quantum emitter separated by a distance $d$ of a dielectric cylinder of relative electric permittivity $\varepsilon$ and radius $R$, coated with graphene. {(b)} Dispersion relation for all the plasmonic modes supported with free-space oscillating light of frequency $\hbar\omega < 200$ meV. Each color is associated with a GCW of a different radius, while the linestyle characterize the order of the mode. {(c)} 
The ratio between the real and imaginary parts of $k_p$ for each plasmonic mode as a function of frequency. In both plots we choose silicon as the dielectric medium, which has permittivity $\varepsilon = 11.68$ in the frequency range considered. Also, $R = \{20, 30, 42\}$ nm (blue, purple, and green lines, respectively), and the Fermi energy is $E_F = 1$ eV.}
\label{fig3}
\end{center}
\end{figure*}
The quantum efficiency can be interpreted as the probability of emission through the chosen pathway against competition with one-photon emission decay channels. To better understand the impact of the SWCNT in the TPSE, we must mention that the quantum efficiency in free-space is of the order of $10^{-8}$\% for the two-photon $6s \rightarrow 5s$ transition in hydrogen. If the emitter is placed in the vicinities of a SWCNT, the quantum efficiency can reach values of $\sim 1$\% in the near-field ($d \lesssim 20$ nm), which is about the same value reported for an emitter close to an extended graphene monolayer~\cite{rivera2016}. Hence, for a quantum emitter near a SWCNT, one expects to observe TPSE decay with a mean lifetime $\sim 20$ ns, which contrasts with a TPSE mean lifetime of a few days for decay in free-space. It is important to emphasize that this significant increase in the probability of second-order decay is achieved through the emission of plasmons, not photons, as in the case of an atom in free-space.  However, since the one-dimensional geometry of a SWCNT limits the possible directions in which the plasmons are allowed to propagate, it may be possible to achieve higher conversion rates of the entangled plasmons into photons by scattering processes due to the presence of defects. Finally, we notice in the plot that the TPSE efficiency keeps above $0.01$ \% until $d \sim 100$ nm.
This robustness of the efficiency of emission may be of practical interest in situations where one does not have precise control over the distance between the emitter and the nanotube.

\subsection{TPSE in graphene-coated nanowire}

Now we turn our attention to the case of an emitter close to a cylindrical waveguide coated with graphene, as shown in Fig. 3a. Unlike SWCNTs, graphene-coated wires are more stable and do not have strict constraints on their radius~\cite{zeng2016, he2013, zhu2014}. As a consequence, a multitude of entangled plasmonic modes can be excited in the TPSE process in the infrared region. In Fig. 3b we consider a silicon cylinder ($\varepsilon \simeq 11.68$) covered with graphene and plot the dispersion relation for all supported plasmonic modes with frequencies below $300$ meV. For fixed radius, each $m \neq 0$ mode exists only above its minimum excitation frequency given by Eq. \eqref{omega_min}, while the fundamental mode, such as the case of a SWCNT, can be excited at any frequency. The number of modes present in the frequency range depends on the radius of the nanotube (and also on the Fermi energy and the inner dielectric medium), and some of them are degenerate at specific frequencies. This can be identified when two dispersion curves of the same color in the plot cross each other, which can be seen, for instance, for the fundamental and first-order plasmonic modes. We also demonstrate strong light confinement for the non-fundamental plasmonic modes, which increases for higher values of the radius. The fundamental mode, however, presents slightly smaller confinement, which decreases (increases) with the radius for small (high) frequencies with the change of behavior occurring around $\sim 150$ nm. This same reasoning also explains the variations of ${\rm Re} (k_p)/{\rm Im} (k_p)$ versus frequency, as presented in Fig. 3c. We notice that no trade-off exists between light confinement and propagation length with respect to the radius. In contrast to the effect that increasing the chemical potential has on both quantities, larger wires present smaller FOMs and propagation lengths due to the propagation of plasmons around the wire.

\begin{figure}[t!]
\begin{center}
\includegraphics[width=0.98\linewidth]{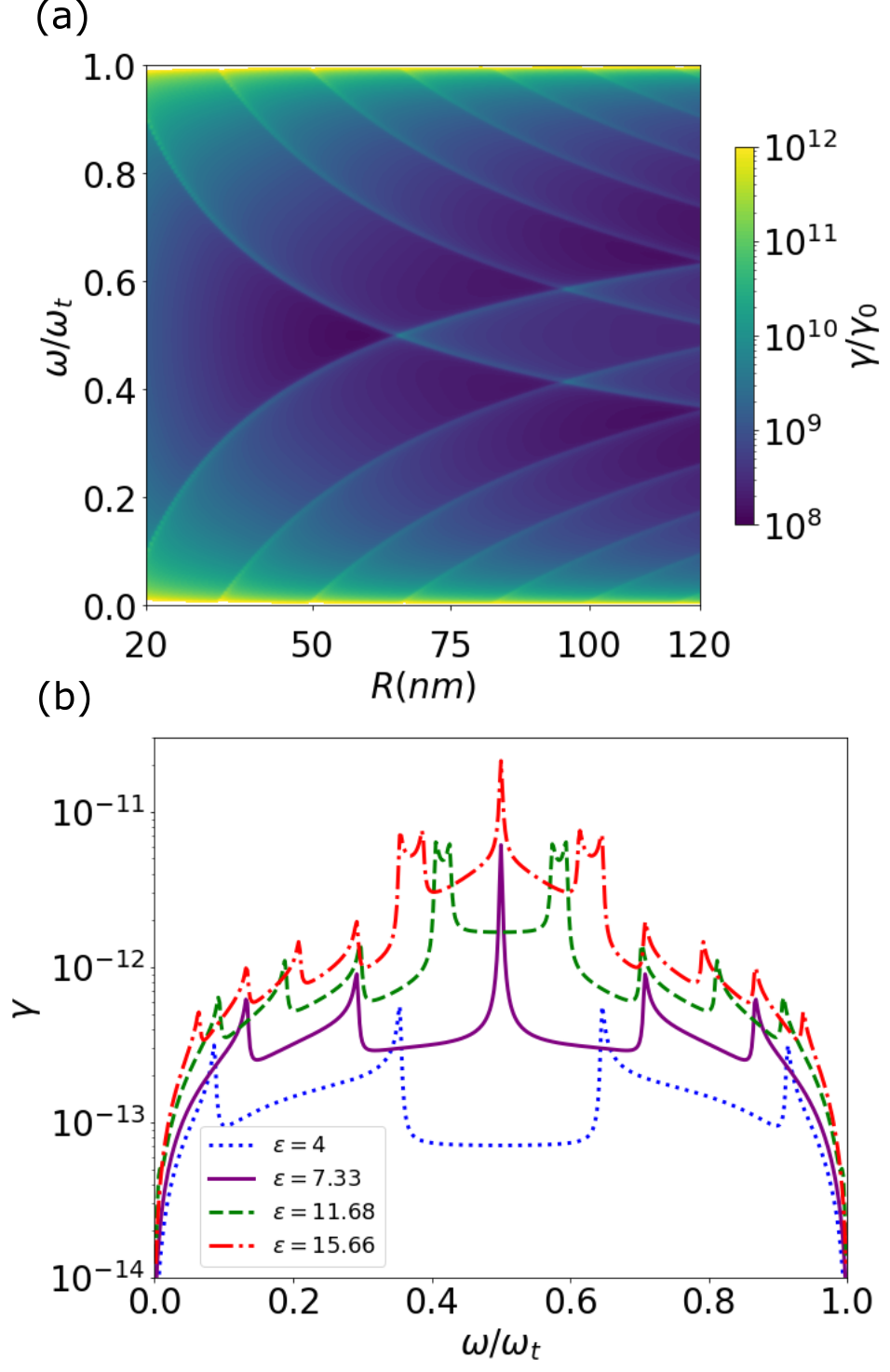}
\caption{(a) TPSE spectral density near a silicon ($\varepsilon = 11.68$) nanowire covered with graphene as a function of the wire radius. The transition frequency considered is $\hbar\omega_t \approx 166$ meV. (b) TPSE spectral density for various dielectric coated nanowires of radius $R = 100$ nm. For $\varepsilon = 7.33$ ($\varepsilon = 15.66$), the minimum excitation frequency of the mode $m = 1$ ($m = 2$) is precisely at $\omega_t/2$, which results in a huge resonance in the middle of the spectrum. In both plots the Fermi energy is $E_F = 1$~eV.}
\label{fig4}
\end{center}
\end{figure}

In Fig. 4a we plot the normalized TPSE spectral density for a transition frequency of $\hbar\omega_t \sim 166$ meV as a function of the normalized frequency of emission (vertical axis) and the silicon GCW radius (horizontal axis). The most flashy aspect of the figure is the multitude of resonances present in the spectrum, which can be accounted by the high degree of degeneracy of the plasmonic modes at their minimum excitation frequencies (as shown in Fig. 3b). Therefore, these resonance curves follow precisely the square root relation given by Eq \eqref{omega_min}. Due to the symmetric aspect of $\gamma(\omega)$, the resonances appears at $\omega_m^{(\text{min})}$ and at $\omega_t - \omega_m^{(\text{min})}$. Hence, cross-talks between modes of different orders exist when their minimum frequencies are complementary, i.e., when $\omega_m^{(\text{min})} + \omega_{m'}^{(\text{min})} = \omega_t$. In the particular case where $\omega_m^{(\text{min})} = \omega_t/2$, a stronger resonance takes place precisely in the middle of the spectrum, where both entangled plasmons are indistinguishable and emitted with the same frequency. These features are very similar to the case of an emitter close to a graphene nanostructure, which was studied in~\cite{muniz2020}. However, we emphasize that GCW plasmons exist in a continuous range of frequencies, in contrast to the well-defined frequencies of graphene nanostructures localized surface plasmons. Therefore, the nature of the resonances supported in the two cases are different. Precisely at the resonance frequencies, the magnitude of the spectral enhancement of an emitter near a GCW is smaller than if it is placed close to a graphene nanostructure. However, the opposite relation holds in the spectral regions between the resonance frequencies since GCW plasmons couple to the emitter at any frequency of emission. In Fig. 4b we fix the radius at $100$ nm and plot the spectral density for different dielectric materials. Since $\omega_m^{(\text{min})} \sim (1 + \varepsilon)^{-1/2}$, higher values of the relative permittivity also increase the number of plasmonic modes contributing to the spectral enhancement. One is able to tailor the dominant modes and frequencies of the emitted entangled plasmons. For $\varepsilon = 4$, only one plasmonic mode besides the fundamental is present on the TPSE spectrum, and $\varepsilon = 7.33$ and $\varepsilon = 15.66$ are chosen such that $\omega_1^{(\text{min})} = \omega_t/2$ and $\omega_2^{(\text{min})} = \omega_t/2$, respectively. In both latter cases, a stronger resonance takes place at half of the transition frequency in comparison to the resonances at other frequencies of emission. The curve of a GCW with $\varepsilon = 11.68$ is a vertical cut of Fig. 4a for $R = 100$ nm but without the normalization by $\gamma_0$.

\begin{figure}[t!]
\begin{center}
\includegraphics[width=0.98\linewidth]{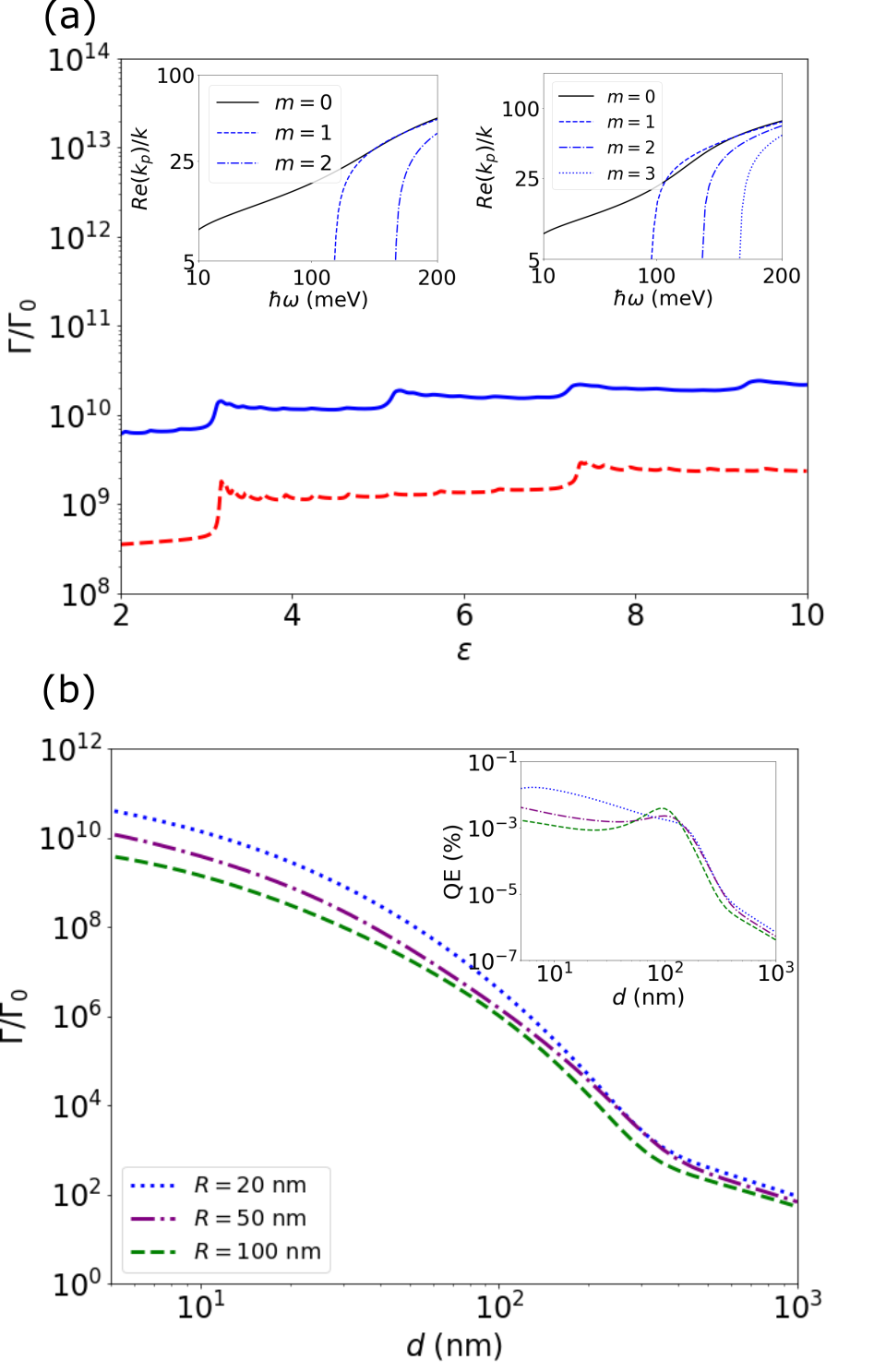}
\caption{(a)  Normalized TPSE rate for the $6s \rightarrow 5s$ transition in hydrogen as a function of the inner dielectric cylinder permittivity. We choose a nanowire of radius $25$ nm and Fermi energies of $0.5$ eV (blue solid curve) and $1$ eV (red dashed curve). Inset: plasmon dispersion relations for $\varepsilon = 3.11$ (left) and $\varepsilon = 5.15$ (right). (b) TPSE rate for the same hydrogen transition as a function of distance. The Fermi energy is equal to $1$ eV. Inset: Quantum efficiency (Eq. \eqref{QE}) as a function of distance for the same parameters of the main plot.}
\label{fig5}
\end{center}
\end{figure}

Despite the increase in $\gamma$ as a function of the permittivity shown in Fig. 4b, the average spectral enhancement varies slowly with $\varepsilon$. As a consequence, the order of magnitude of the TPSE rate, which is the integral of $\gamma(\omega)$, is not substantially affected by changes of the inner dielectric medium for a wide range of values of $\varepsilon$. This allows one to tune the spectrum of emission without significantly affecting the desired high two-plasmon decay rate of the emitter, in contrast to what is feasible in systems that support localized surface plasmons. Indeed, this can be seen in Fig. 5a where we plot the TPSE rate for the hydrogen $6s \rightarrow 5s$ transition as a function of the relative permittivity. Lowering the Fermi energy, however, increases the plasmonic density of states in the whole spectrum of emission, which results in a higher TPSE rate. This can be seen by comparing the solid ($E_F = 0.5$ eV) with the dashed curve ($E_F = 1$ eV). For some specific values of the permittivity, we notice resonances in the emission rate, a direct consequence of the degeneracy between the fundamental and first-order plasmonic modes. In the inset of Fig. 5a, we plot the plasmon dispersion relations for $\varepsilon \sim 3.11$ and $\varepsilon \sim 5.15$, which are the values for the first and second resonances for $E_F = 0.5$ eV. We notice that in contrast to Fig. 3b, where the modes are degenerate at two well-defined frequencies, for these values of $\varepsilon$ the degeneracy occurs in a quasi-continuous range of frequencies. Figure 5b shows the TPSE rate and quantum efficiency as a function of distances for different radii. We identify features similar to those in the emitter-SWCNT system analyzed in Fig. 2. In comparison to that case, we conclude that TPSE efficiency in GCW is more robust to distance variations than in the case of a SWCNT, with a small increase in magnitude right before starting to approach its free-space value. This can be explained by the fact that the one-plasmon SE rates in the denominator of Eq. \eqref{QE} decay to their free-space values more rapidly than the TPSE rate in the numerator. 


\section{Conclusions}
In conclusion, we investigated the spontaneous emission of two plasmons by single quantum emitters in low-dimensional carbon nanomaterials such as single-walled carbon nanotubes and graphene-coated nanowires. We have shown that SWCNTs are a suitable material platform to increase the emission rate by more than twelve orders of magnitude concerning the rate in free-space and with average lifetimes of the order of a few dozens of nanoseconds. Such impressive enhancements are possible due to the large confinement of the fundamental plasmonic mode in a broad frequency range. In order to extend our investigation to larger plasmonic nanowires, we considered the case of a dielectric cylinder coated with a graphene monolayer. We demonstrate the role of different plasmonic modes supported by the nanowire in TPSE, which results in a rich, tunable broadband spectrum of emission, with sharp resonances that precisely occur at the plasmons' minimum excitation frequencies. We concluded that GCWs enable bespoke tailoring of the spectral lineshapes while significantly minimizing the emitter's lifetime. Our results pave the way for applying guided plasmons in one-dimensional carbon nanostructures to enhance and tailor TPSE, extending the possibilities of generating entangled plasmonic and photonic excitations by means of non-linear atomic transitions.

\section{Acknowledgements}
Y.M., F.A.P., and C.F. acknowledge funding by the Coordena\c c\~ ao de Aperfei\c coamento de Pessoal de N\'ivel Superior (CAPES). P.P.A., F.A.P., and C.F. acknowledge Fundação de Amparo à Pesquisa do Estado do Rio de Janeiro (FAPERJ). F.A.P. and C.F. acknowledge Conselho Nacional de Desenvolvimento Cient\'ifico e Tecnol\'ogico (CNPq) - research grant 310365/2018-0. L.M.M. acknowledges Project PID2020-115221GB-C41 financed by MCIN/AEI/10.13039/501100011033, the Ministry of Science and Higher Education of the Russian Federation (Agreement No. 075-15-2021-606) and Aragon Government through Project Q- 503 MAD. W.K.-K. acknowledges funding from the Laboratory Directed Research and Development program of Los Alamos National Laboratory under projects number 20190574ECR and 20220228ER.
%

\end{document}